\documentclass[aps,pre,twocolumn,floats,showpacs,superscriptaddress]{revtex4}
\usepackage{epsfig}
\usepackage{amsmath}

\newcommand{\equ}[1]{(\protect\ref{#1})}
\newcommand{\avk}{\left< k \right>}
\newcommand{\fluck}{\left< k^2 \right>}
 
\begin{document}

\title{Immunization of complex networks}

\author{Romualdo Pastor-Satorras} 

\affiliation{Departament de F{\'\i}sica i Enginyeria Nuclear, Universitat
  Polit{\`e}cnica de Catalunya, Campus Nord, M\`{o}dul B4, 08034
  Barcelona, Spain}

\author{Alessandro Vespignani}

\affiliation{The Abdus Salam International Centre for Theoretical
  Physics (ICTP), P.O. Box 586, 34100 Trieste, Italy}

\date{\today}

\begin{abstract}
  Complex networks such as the sexual partnership web or the Internet
  often show a high degree of redundancy and heterogeneity in their
  connectivity properties. This peculiar connectivity provides an
  ideal environment for the spreading of infective agents.  Here we
  show that the random uniform immunization of individuals does not
  lead to the eradication of infections in all complex networks.
  Namely, networks with scale-free properties do not acquire global
  immunity from major epidemic outbreaks even in the presence of
  unrealistically high densities of randomly immunized individuals.
  The absence of any critical immunization threshold is due to the
  unbounded connectivity fluctuations of scale-free networks.
  Successful immunization strategies can be developed only by taking
  into account the inhomogeneous connectivity properties of scale-free
  networks.  In particular, {\em targeted} immunization schemes, based
  on the nodes' connectivity hierarchy, sharply lower the network's
  vulnerability to epidemic attacks.
\end{abstract}

\maketitle

\section{Introduction}

The relevance of spatial and other kinds of heterogeneity in the
design of immunization strategies has been widely addressed in the
epidemic modeling of infectious diseases
\cite{anderson92,epidemics}.  In particular, it has been pointed
out that population inhomogeneities can substantially enhance the
spread of diseases, making them harder to eradicate and calling for
specific immunization strategies. This issue assumes the greatest
importance in a wide range of natural interconnected systems such as
food-webs, communication and social networks, metabolic and neural
systems~\cite{amaral,strog01}.  The complexity of these networks
resides in the small average path lengths among any two nodes
(small-world property), along with a large degree of local clustering.
In other words, some special nodes of the structure develop a larger
probability to establish connections pointing to other nodes.  This
feature has dramatic consequences in the topology of scale-free (SF)
networks \cite{barab99,falou99,mendes99} which exhibit a power-law
distribution
\begin{equation}
  P(k)\sim k^{-\gamma}
\end{equation}
for the probability that any node has $k$ connections to other nodes.
For exponents in the range $2 < \gamma \leq 3$, this connectivity distribution
implies that, for large network sizes, the nodes have a statistically
significant probability of having a very large number of connections
compared to the average connectivity, $\left<k\right>$.  This feature
contrasts with what is found for homogeneous networks (local or
nonlocal) in which each node has approximately the same number of
links, $k\simeq \left<k\right>$~\cite{erdos60,watts98}. The extreme
heterogeneity of SF networks finds the most stunning examples in two
artificial systems, the World-Wide-Web (WWW) \cite{barab99,www99} and
the Internet \cite{falou99,calda00,alexei}.  Along with these
technological networks, it has also been pointed out that sexual
partnership networks are often extremely heterogeneous
\cite{het84,anderson92,may87}, and it has been recently observed that
the network of sexual human contacts possesses a well-defined
scale-free nature \cite{amaral01}.

In homogeneous networks, an epidemic occurs only if the rate of
infection of ``healthy'' individuals connected to infected ones
exceeds the so-called {\em epidemic threshold}; in other words, if the
disease cannot transmit itself faster than the time of cure, it dies
out~\cite{anderson92,epidemics}. In heterogeneous networks, on the
other hand, it is well-known that the epidemic threshold decreases
with the standard deviation of the connectivity
distribution~\cite{anderson92}.  This feature is paradoxically
amplified in scale-free networks which have diverging connectivity
fluctuations. In fact, as it was first noted in
Refs.~\cite{pv01a,pv01b}, epidemic processes in SF networks do not
possess, in the limit of an infinite network, an epidemic threshold
below which diseases cannot produce a major epidemic outbreak or the
inset of an endemic state. SF networks are therefore prone to the
spreading and the persistence of infections, whatever virulence the
infective agent might possess.

In view of this weakness, it becomes a major task to find optimal
immunization strategies oriented to minimize the risk of epidemic
outbreaks on SF networks, task with immediate practical and economical
implications. This paper presents a parallel comparison of the effect
of different immunization schemes in the case of two different complex
networks: the Watts-Strogatz model \cite{watts98} and the Barab\'{a}si
and Albert model \cite{barab99}.  The first is a homogeneous network
exhibiting small-world properties, while the second one is the
prototype example of SF network.  By studying the
susceptible-infected-susceptible model \cite{epidemics} in presence of
progressively greater immunization rates, we find that uniformly
applied immunization strategies are effective only in complex networks
with bounded connectivity fluctuations.  On the contrary, in SF
networks the infection is not eradicated even in the presence of an
unrealistically high fraction of immunized individuals.  Actually, SF
systems do not have any critical fraction of immunized individuals and
only the total immunization of the network achieves the infection's
eradication.  In order to overcome these difficulties we define
optimal immunization strategies that rely on the particular SF
structure of the network.  The developed strategies allow us to
achieve the total protection of the network even for extremely low
fractions of successfully immunized individuals.

\section{The model}

In order to estimate the effect of an increasing density of immune
individuals in complex networks, we will investigate the standard
susceptible-infected-susceptible (SIS) model~\cite{epidemics}.  This
model relies on a coarse-grained description of individuals in the
population.  Namely, each node of the graph represents an individual
and each link is a connection along which the infection can spread.
Each susceptible (healthy) node is infected with rate $\nu$ if it is
connected to one or more infected nodes.  Infected nodes are cured and
become again susceptible with rate $\delta$, defining an effective
spreading rate $\lambda=\nu/\delta$ (without lack of generality, we
set $\delta=1$).  The SIS model does not take into account the
possibility of individuals' removal due to death or acquired
immunization \cite{epidemics}, and thus individuals run stochastically
through the cycle susceptible $\to$ infected $\to$ susceptible.  This
model is generally used to study infections leading to endemic states
with a stationary average density of infected individuals.

\subsection{Homogeneous complex networks}

A wide class of network models \cite{erdos60,watts98} have
exponentially bounded connectivity fluctuations. A paradigmatic
example of this kind of networks that has recently attracted a great
deal of attention is the Watts-Strogatz (WS) model \cite{watts98},
which is constructed as follows: The starting point is a ring with $N$
nodes, in which each node is symmetrically connected with its $2K$
nearest neighbors.  Then, for every node each link connected to a
clockwise neighbor (thus $K$ links for each node) is kept as originating 
from the original node and rewired to a randomly chosen target node with
probability $p$. This procedure
generates a random graph with a connectivity distributed exponentially
for large $k$, and an average connectivity $\left<k \right> = 2 K$.
It is worth remarking that even in the case $p=1$ the network keeps the 
memory of the construction algorithm and is not equivalent to 
a random graph. In fact, by definition each node emanates at least the
$K$ links which have been rewired from the clockwise neighbors to 
randomly chosen nodes;
a property that affects also the clustering properties of the graph 
(for details see Ref.~\cite{barrat00}).
  
For the class of exponentially bounded networks, 
one can generally consider that each node has roughly the same 
number of links, $k\simeq \left< k
\right>$, and therefore we can consider them as fairly homogeneous 
in their connectivity properties.
At a mean-field level, the equation
describing the time evolution of the average density of infected
individuals $\rho(t)$ (prevalence) is
\begin{equation}
\frac{d  \rho(t)}{d t} = -\rho(t) +\lambda \left< k \right> 
\rho(t) \left[ 1-\rho(t) \right].
\label{eq:ws}
\end{equation}
The mean-field character of this equation stems from the fact that we
have neglected the density correlations among the different nodes,
independently of their respective connectivities.  The first term on
the r.h.s. in Eq.~(\ref{eq:ws}) considers infected nodes becoming
healthy with unit rate. The second term represents the average density
of newly infected nodes generated by each active node. This is
proportional to the infection spreading rate $\lambda$, the number of
links emanating from each node $k\simeq \left< k
\right>$, and the probability that a given link
points to a healthy node, $\left[ 1-\rho(t) \right]$.
After imposing the stationary condition $d\rho(t) / dt =0$, 
the most significant and general result is the
existence of a nonzero epidemic threshold $\lambda_c=\left< k
\right>^{-1}$ \cite{epidemics} such that
\begin{eqnarray}
  \rho &=& 0 \;\qquad\qquad \mbox{\rm if $\lambda< \lambda_c$}, \\
  \rho &\sim& \lambda-\lambda_c \qquad \mbox{\rm if
  $\lambda\geq\lambda_c$} \label{eq:meanfield}.
\end{eqnarray}
In other words, if the value of $\lambda$ is above the threshold,
$\lambda\geq \lambda_c$, the infection spreads and becomes endemic.
Below it, $\lambda <\lambda_c$, the infection dies out exponentially
fast.  The existence of an epidemic threshold is a general results in
epidemic modeling, present also in different models such as the
susceptible-infected-removed (SIR) model \cite{epidemics}. In analogy
with critical phenomena \cite{marro99}, this kind of behavior can be
identified as an absorbing-state phase transition, in which $\rho$
plays the role of the order parameter in the phase transition and
$\lambda$ is the tuning parameter, recovering the usual mean-field
behavior \cite{marro99}.

\subsection{Scale-free networks}

This standard framework is radically changed in the class of SF
networks \cite{pv01a,pv01b}, in which the probability distribution
that a node has $k$ connections has the form $P(k) \sim k^{-\gamma}$
and the connectivity fluctuations, $\left< k^2 \right>$, diverge in
infinite networks for any value $2<\gamma\leq 3$. The paradigmatic
example of SF network is the Barab\'{a}si and Albert (BA) model
\cite{barab99}.  The construction of the BA graph starts from a small
number $m_0$ of disconnected nodes; every time step a new vertex is
added, with $m$ links that are connected to an old node $i$ with
probability $\Pi(k_i) = k_i / \sum_j k_j$, where $k_i$ is the
connectivity of the $i$-th node.  After iterating this scheme a
sufficient number of times, we obtain a network composed by $N$ nodes
with connectivity distribution $P(k) \sim k^{-3}$ and average
connectivity $\left<k \right> = 2 m$.  For this class of graphs, the
absence of a characteristic scale for the connectivity makes highly
connected nodes statistically significant, and induces strong
fluctuations in the connectivity distribution which cannot be
neglected.  In order to take into account these fluctuations, we have
to relax the homogeneity assumption used for homogeneous networks, and
consider the relative density $\rho_k(t)$ of infected nodes with given
connectivity $k$; i.e., the probability that a node with $k$ links is
infected. The dynamical mean-field equations can thus be written as
\cite{pv01a,pv01b}
\begin{equation}
  \frac{ d \rho_k(t)}{d t} = -\rho_k(t) +\lambda k \left[
  1-\rho_k(t) \right] \Theta(\rho(t)), 
\label{mfk}
\end{equation}
where also in this case we have considered a unity recovery rate.
The creation term considers the probability that a node with $k$ links
is healthy $[1-\rho_k(t)]$ and gets the infection via a
connected node.  The probability of this last event is proportional
to the infection rate $\lambda$, the number of connections $k$, and
the probability $\Theta(\rho(t))$ that any given link points to an
infected node. The probability that a link points to 
a node with $s$ links is
proportional to $sP(s)$.  In other words, a randomly chosen link is
more likely to be connected to an infected node with high
connectivity, yielding
\begin{equation}
  \Theta(\rho(t))= \frac{\sum_k kP(k)\rho_k(t)}{\sum_s sP(s)},
  \label{first}
\end{equation}
where $\sum_s sP(s)$ is identical to $\left< k \right>$ by definition.
In the stationary state [$d \rho_k(t)/ dt =0$], Eq.~(\ref{mfk})
yields the following  infected node density form
\begin{equation}
  \rho_k= \frac{\lambda k \Theta}{1 + \lambda k \Theta}.
\end{equation}
By inserting the above  expression for $\rho_k$ in Eq.~(\ref{first}),
we obtain the self-consistency equation 
\begin{equation}
  \Theta = \frac{1}{\avk}  \sum_k k P(k)  \frac{\lambda k 
\Theta}{1 + \lambda k \Theta},
\label{cons}
\end{equation}
where $\Theta$ is now a function of $\lambda$
alone~\cite{pv01a,pv01b}.  The solution $\Theta=0$ is always
satisfying the consistency equation. A non-zero stationary prevalence
($\rho_k\neq 0$) is obtained when the r.h.s. and the l.h.s. of
Eq.~(\ref{cons}), expressed as function of $\Theta$, cross in the
interval $0<\Theta\leq 1$, allowing a nontrivial solution.  It is easy
to realize that this corresponds to the inequality
\begin{equation}
  \frac{d}{d \Theta } \left. \left( \frac{1}{\avk}  
      \sum_k k P(k)  \frac{\lambda k
        \Theta}{1 + \lambda k \Theta} 
    \right) \right|_{\Theta =0} \geq 1
  \label{eq:critpunt}
\end{equation}
being satisfied. The value of $\lambda$ yielding the equality in
Eq.~\equ{eq:critpunt} defines the critical epidemic threshold
$\lambda_c$, that is given by
\begin{equation}
  \frac{\sum_k k P(k)  \lambda_c k}{\avk}   =  
\frac{\fluck}{\avk}  \lambda_c= 1
  \quad \Rightarrow \quad  \lambda_c = \frac{\avk}{\fluck}.
\label{thr}
\end{equation}
This results implies that in SF networks with connectivity exponent
$2<\gamma\leq 3$, for which $\fluck\to\infty$, we have $\lambda_c=0$.
This fact implies in turn that for any positive value of $\lambda$ the
infection can pervade the system with a finite prevalence, in a
sufficiently large network~\cite{pv01a,pv01b}.  For small $\lambda$ it
is possible to solve explicitly Eq.~(\ref{cons}) for SF networks and
calculate the prevalence in the endemic state as $\rho=\sum_k
P(k)\rho_k$ as shown in  Refs.~\cite{pv01a,pv01b}.  Calculations can
be carried out by using the continuous $k$ approximation, valid for
large $k$ \cite{barab992}, that assumes $\left<
  k^n\right>=\int_m^{\infty} k^nP(k)dk$, where $m$ is the minimum
number of connections of any node and $P(k)$ is a properly defined
probability density of connections.  For the particular case of the BA
network we have $P(k)=2m^2k^{-3}$ \cite{barab99}, that in the limit of
an infinitely large network yields the prevalence \cite{pv01a,pv01b}
\begin{equation}
\rho\simeq 2 \exp(-1/m \lambda).
\label{op}
\end{equation}

Obviously, $\fluck$ assumes a bounded value in finite size networks,
defining an effective threshold $\lambda_c(N) > 0$ due to finite size
effects, as customarily encountered in nonequilibrium phase
transitions \cite{marro99}.  This epidemic threshold, however, is not
an {\em intrinsic} quantity as in exponential networks and it is
vanishing for increasing network sizes; i.e. in the thermodynamic
limit.  Since real networks have always a finite size, however, it is
interesting to calculate how the epidemic threshold scales with the
system size \cite{lloydsir}. By considering the continuous $k$
approximation, it is possible to calculate the finite size
distribution moments as $\left< k^n \right>=\int_m^{k_c} k^nP(k)dk$,
where $k_c$ is the largest connectivity present in the finite network.
For networks composed by $N$ nodes, $k_c$ is obviously an increasing
function of $N$.  In the particular case of the BA model, we readily
obtain $\left< k \right>\simeq 2m$ and $\left< k^2 \right>\simeq
2m^2\ln(k_c/m)$ as $k_c\to\infty$. Substituting this values in the
Eq.~\equ{thr} we obtain a threshold
$\lambda_c\simeq(m\ln(k_c/m))^{-1}$.  In order to find the size
dependence of $\lambda_c$, we have to relate the maximum connectivity
$k_c$ with the network size $N$. This relation is given by $k_c \simeq
mN^{1/2}$ \cite{barab992,dorogorev}, yielding finally a threshold
\begin{equation}
  \lambda_c(N) = \frac{\avk}{\fluck} \sim \frac{1}{\ln(N)}.
  \label{eq:lambdaN}
\end{equation}
This result can be generalized to SF networks with an arbitrary
connectivity distribution, which show an epidemic threshold vanishing
as a power law behavior in $N$ with an exponent depending on the
connectivity exponent $\gamma$ \cite{pvbrief}.

\section{Uniform immunization strategy}

The simplest immunization procedure one can consider consists in the
random introduction of immune individuals in the
population~\cite{anderson92}, in order to get a uniform immunization
density.  Immune nodes cannot become infected and thus do not transmit
the infection to their neighbors. In this case, for a fixed spreading
rate $\lambda$, the relevant control parameter is the immunity $g$,
defined as the fraction of immune nodes present in the network.  
At the mean-field level, the presence of uniform
immunity will effectively reduce the spreading rate $\lambda$ by a
factor $(1-g)$; i.e. the probability of finding and infecting a
susceptible and non-immune node.  By substituting
$\lambda\to\lambda(1-g)$ in Eqs.~\equ{eq:ws} and \equ{mfk} we obtain
the prevalence behavior for progressively larger immunization rates.

In homogeneous networks, such as the WS model, it is easy to show that
in the case of a constant $\lambda$, the stationary  prevalence 
obtained from Eq.~(\ref{eq:ws}) is given by
\begin{eqnarray}
  \rho &=& 0 \;\qquad\qquad \mbox{\rm if $g>g_c$}~, \\
  \rho &\sim& g_c-g  \qquad \mbox{\rm if
  $g\leq g_c$}~. 
\label{eq:imm_ws1}
\end{eqnarray}
Here, $g_c$ is the  critical immunization value
above which the density of infected individuals in the stationary
state is null and depends on $\lambda$ as
\begin{equation}
  g_c=\frac{\lambda-\lambda_c}{\lambda}.
  \label{eq:gcWS}
\end{equation}
Thus, the critical immunization which achieves eradication is
related to the spreading rate and the epidemic threshold of the
infection. Eq.~(\ref{eq:gcWS}) is obviously valid only for 
$\lambda>\lambda_c$, and it
implies that the critical immunization allowing the complete
protection of the network (null prevalence) is increasing with the
spreading rate $\lambda$.

On the contrary, uniform immunization strategies on SF networks are
totally ineffective.  The presence of immunization depresses the
infection's prevalence too slowly, and it is impossible to find any
critical fraction of immunized individuals that ensures the infection
eradication. The absence of an epidemic threshold ($\lambda_c=0$) in
the thermodynamic limit implies that whatever rescaling
$\lambda\to\lambda(1-g)$ of the spreading rate does not eradicate the
infection except the case $g=1$.  In fact, by using Eq.~(\ref{thr}) we
have that the immunization threshold is given by
\begin{equation}
1-g_c=\frac{1}{\lambda}\frac{\avk}{\fluck}.
\label{eq:gcdef}
\end{equation}
In SF networks with $\fluck\to \infty$ only a complete immunization of
the network (i.e., $g_c=1)$ ensures an infection-free stationary
state. The fact that uniform immunization strategies are less
effective has been noted in several cases of spatial heterogeneity
\cite{anderson92}. In SF networks we face a limiting case due to the
extremely high (virtually infinite) heterogeneity in connectivity
properties. Also in this case finite networks present an effective
threshold $g_c(N)$ depending on the number of nodes $N$. As for the
epidemic threshold, however, we are not in presence of an intrinsic
quantity and we have that $g_c(N)\to 1$ in the thermodynamic limit
$N\to\infty$. In the case of the BA model, inserting the expression
\equ{eq:lambdaN} into Eq.~\equ{eq:gcdef}, we observe that the
immunization threshold scales as
\begin{equation}
1- g_c(N) \sim \frac{1}{\lambda\ln(N)}. 
\end{equation}
Also in this case it is possible to generalize this result for
arbitrary connectivity exponents $\gamma$ \cite {pvbrief}.

In order to provide further support to the present mean-field
(sometimes called the deterministic approximation) description, we
study by means of numerical simulations the behavior of the SIS model
on the WS and the BA networks.  In these systems, because of the
nonlocal connectivity, mean-field predictions are expected to
correctly depict the model's behavior.  In the present work we
consider the parameters $K=3$ and maximal disorder $p=1$ for the WS
network, and $m_0=5$ and $m=3$ in the case of the BA network.

\begin{figure*}[t]
  \centerline{\epsfig{file=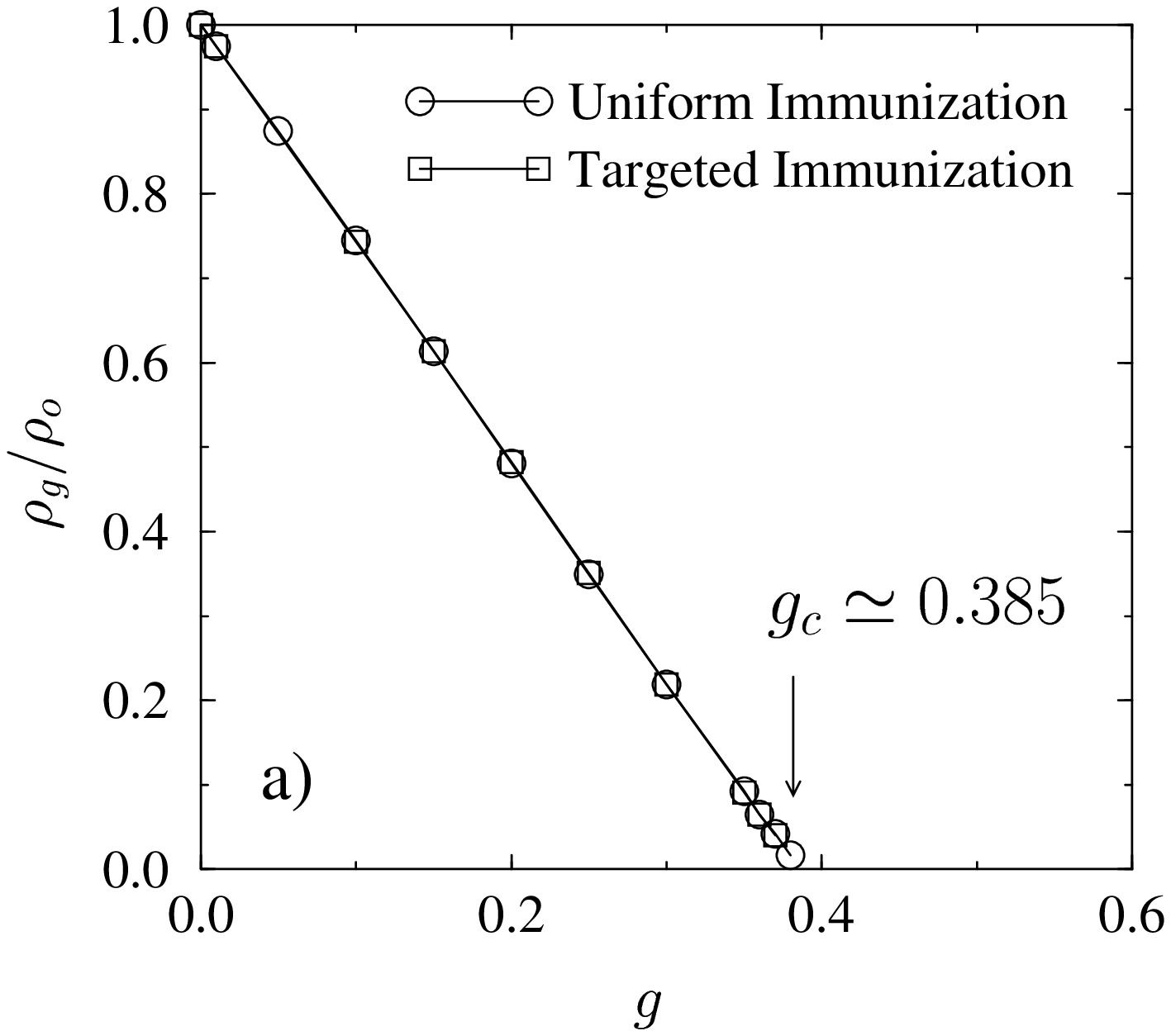, width=8cm} 
    \epsfig{file=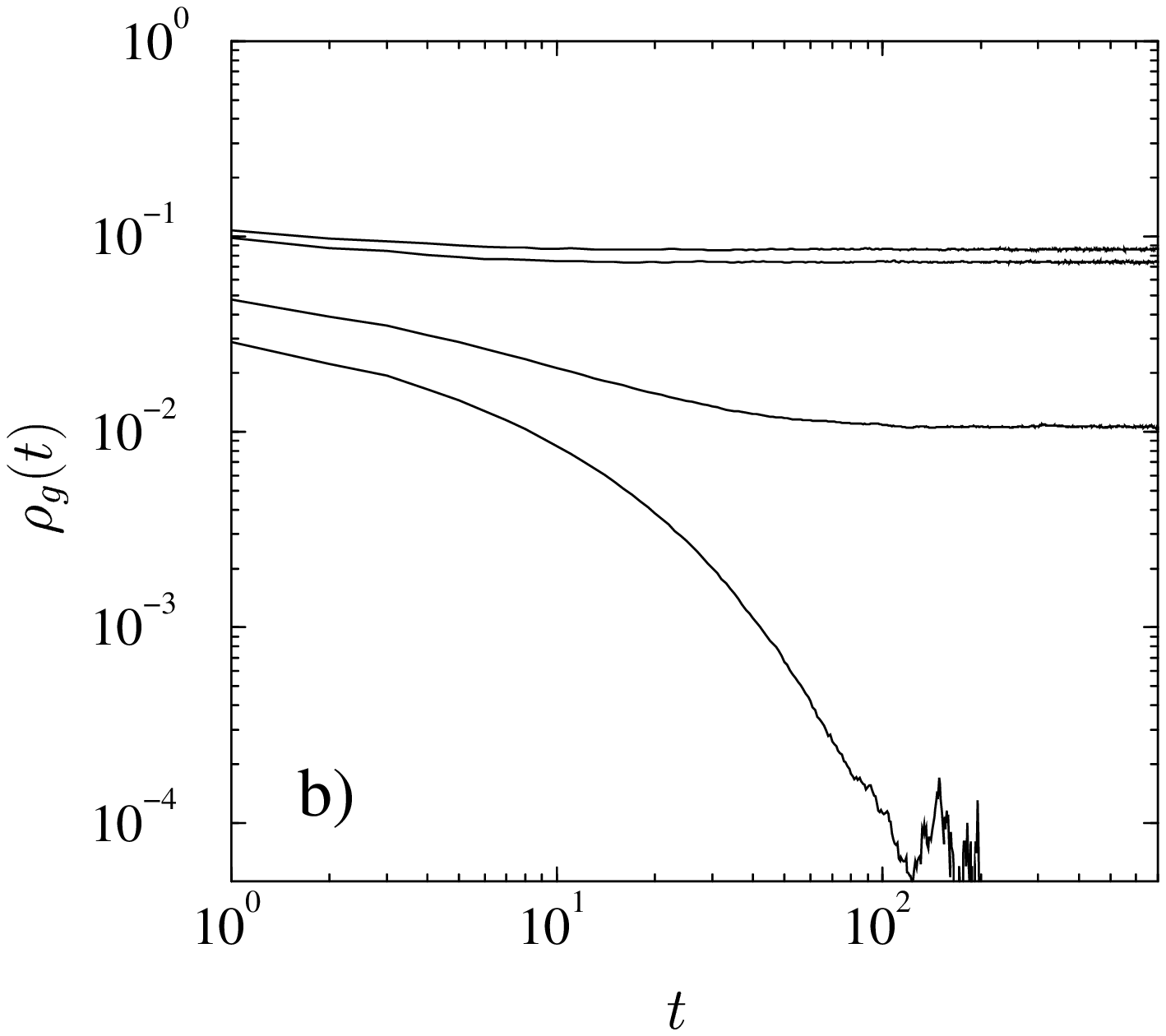, width=8cm}}

  \caption{a) Reduced prevalence $\rho_g/ \rho_0$ from computer
    simulations of the SIS model in the WS network with uniform and
    targeted immunization at a fixed spreading rate $\lambda=0.25$.
    Extrapolation of the linear behavior of $\rho_g$ for the largest
    immunization values yields an estimate of the critical immunity
    $g_c \simeq 0.385$.%
    ~b) Typical plots of $\rho_g(t)$ as a function of time, averaged over
    $100$ starting configurations, for the SIS model in WS networks
    with uniform immunization, for different values of $g$.  From top
    to bottom: $g= 0.1$, $0.14$, $0.35$, and $0.43$.  For the last
    value of $g$ (above the critical immunization) all runs die,
    independently of the network size $N$.}

\end{figure*}

In the presence of uniform immunization, we can study the system by
looking at the infection's prevalence in the stationary regime
(endemic state) as a function of the immunity $g$.  The uniform
immunization is implemented by randomly selecting and immunizing $gN$
nodes on a network of fixed size $N$. Our simulations are implemented
at a fixed spreading rate $\lambda=0.25$. The number of nodes range
from $N=10^4$ to $N=10^6$. We analyze the stationary properties of the
density of infected nodes $\rho_g$ (the infection prevalence) for
different values of the immunization $g$.  Initially we infect half of
the susceptible nodes in the network, and iterate the rules of
the SIS model with parallel updating.  The prevalence is computed
averaging over at least $100$ different starting configurations,
performed on at least $10$ different realizations of the network.  In
Fig. 1(a), we show the behavior of the reduced prevalence $\rho_g/
\rho_0$ (where $\rho_0$ is the prevalence without immunization) as a
function of the uniform immunization $g$ in the WS network.  We
observe that the prevalence of infected nodes decays drastically for
increasing immunization densities (see Fig. 1(b)). In particular, we
observe the presence of a sharp immunization threshold
$g_c\simeq0.385$, in fair agreement with the estimate $g_c \simeq
0.36$ from Eq.~\equ{eq:gcWS} with the values $\lambda=0.25$ and the
estimate $\lambda_c \simeq 0.16$ from Ref.~\cite{pv01b}.  In the
biological case, this effect motivates the use of global vaccination
campaigns in homogeneous populations in order to reach a density of
immune individuals that secures from major outbreaks or endemic
states.  On the contrary, the results for the SF network, depicted in
Fig. 2(a), show a strikingly different behavior.  Namely, the density
of infected individuals decays slowly with increasing immunization,
and it would be null only for the complete immunization of the whole
network ($g=1$).  Specifically, it follows from Eq.~(\ref{op}) that
the SIS model on the BA network shows for $g\simeq 1$ and any
$\lambda$ the prevalence
\begin{equation}
  \rho_g\simeq 2 \exp(-1/m \lambda (1-g))
\end{equation}
We
have checked this prediction in Fig.~2(b). In other words, the
infection always reaches an endemic state if the network size 
is enough large (see Fig.~3(a)). This points
out the absence of an immunization threshold; SF networks are weak in
face of infections, also after massive uniform vaccination campaigns.

\begin{figure*}[t] 
  \centerline{\epsfig{file=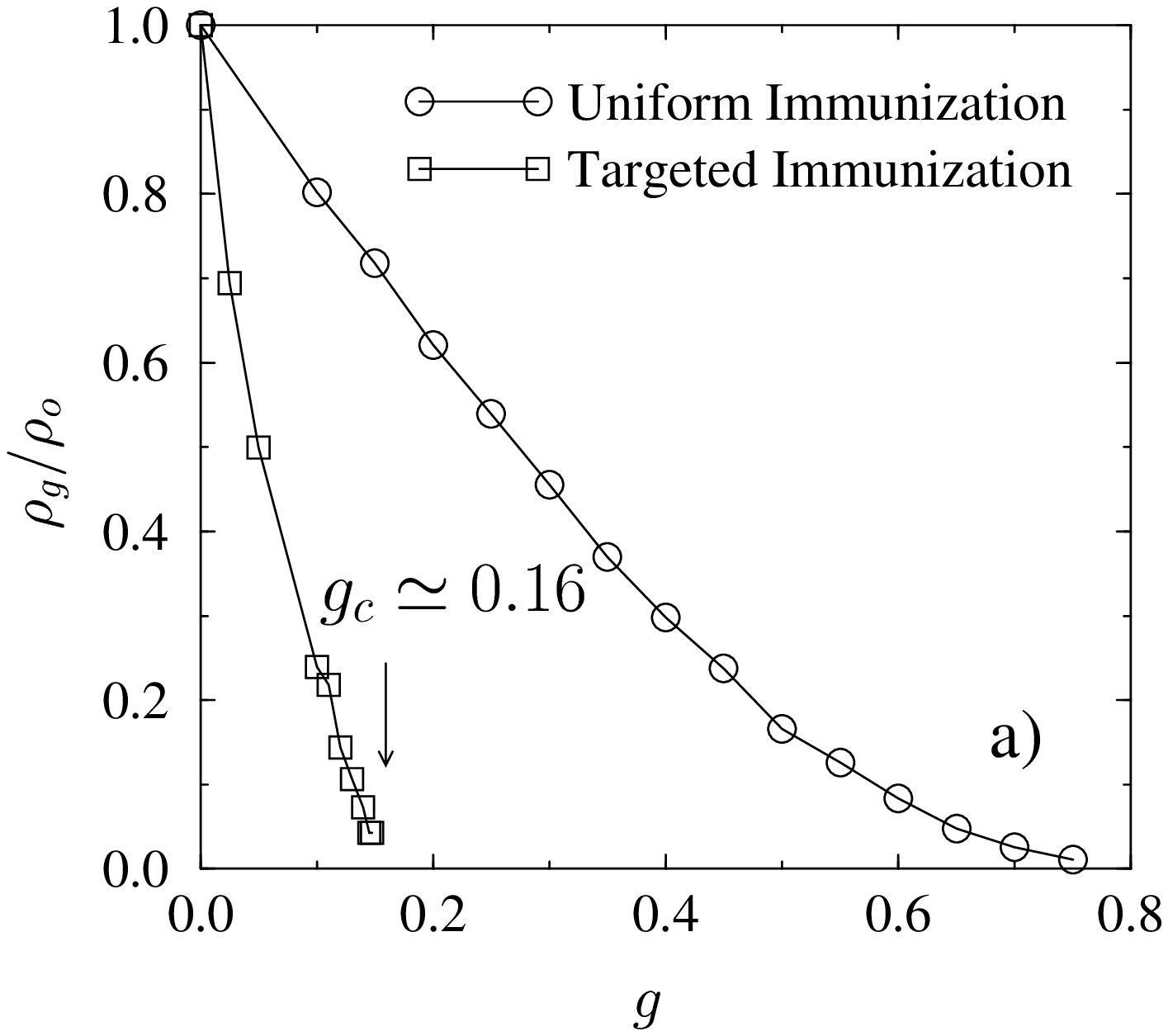, width=8cm}
    \epsfig{file=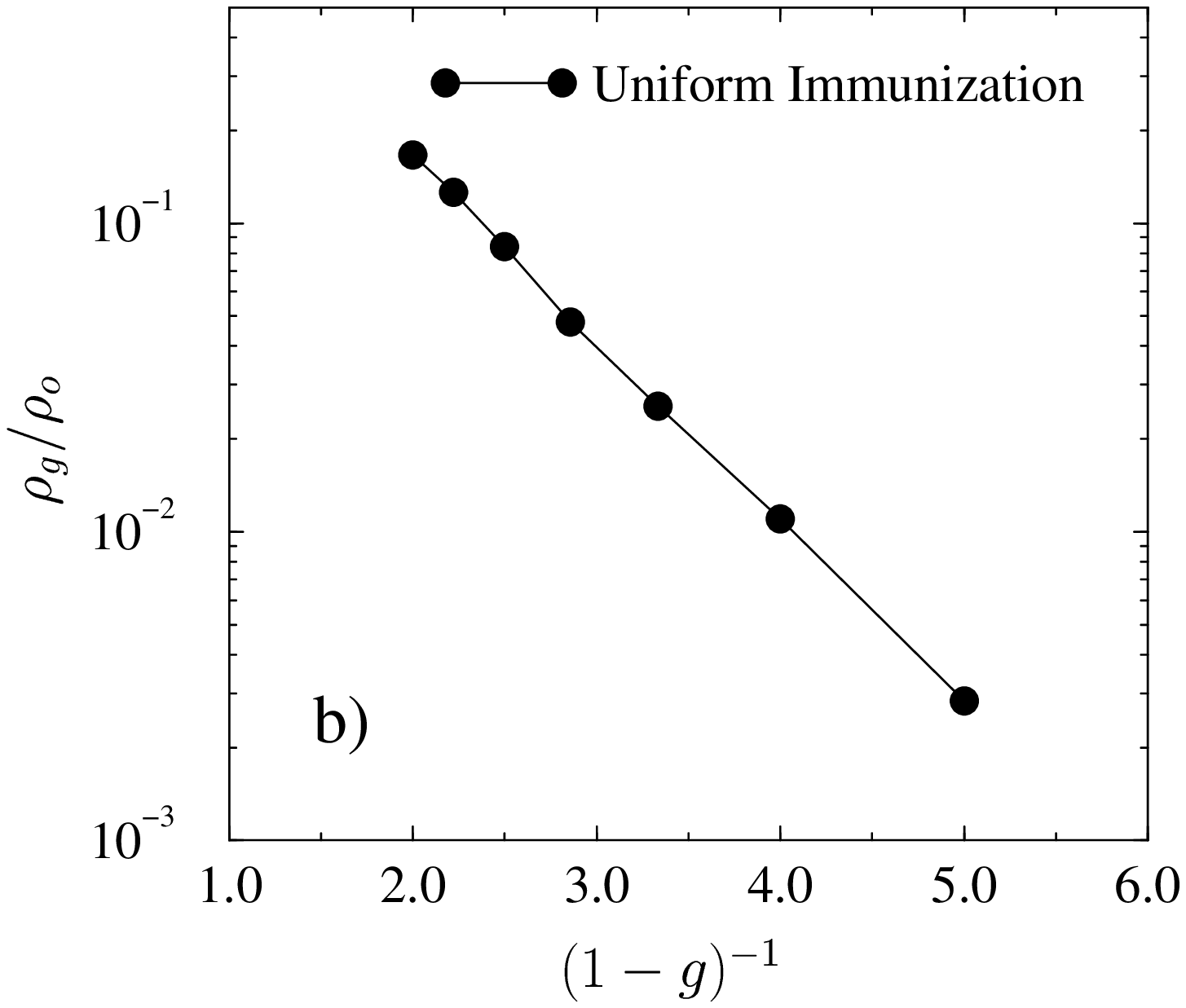, width=8cm}} 
  \caption{a) Reduced prevalence $\rho_g/ \rho_0$ from computer
    simulations of the SIS model in the BA network with uniform and
    targeted immunization, at a fixed spreading rate $\lambda=0.25$.
    A linear extrapolation from the largest values of $g$ yields an
    estimate of the threshold $g_c \simeq 0.16$ in BA networks with
    targeted immunization. %
    ~b) Check of the predicted functional dependence $\rho_g\sim
    \exp(-1/m\lambda(1-g))$ for the SIS model in the BA network with uniform
    immunization. }

\end{figure*}

\section{Optimized immunization strategies}

When fighting an epidemic in an heterogeneous population with a
uniform vaccination scheme, it is necessary to vaccinate a fraction of
the population larger than the estimate given by a simple
(homogeneous) assumption \cite{anderson92}. In this case, it can be
proved \cite{anderson92} that {\em optimal} vaccination programs can
eradicate the disease vaccinating a smaller number of individuals.  SF
networks can be considered as a limiting case of heterogeneous systems
and it is natural to look for specifically devised immunization
strategies.

\subsection{Proportional immunization}
A straightforward way to reintroduce an intrinsic immunization
threshold in SF networks consists in using different fractions of
immunized individuals according to their connectivity. Let us define
$g_k$ as the fraction of immune individuals with a given connectivity
$k$. If we impose the condition
\begin{equation}
\tilde{\lambda}\equiv\lambda k (1-g_k)= {\rm const.},
\end{equation}
we observe that Eqs.~(\ref{mfk}) become identical and decoupled,
defining effectively an homogeneous system. The density of infected
individuals is the same for all connectivities $k$, and an epidemic
threshold $\tilde{\lambda}_c=1$ is reintroduced in the system.  This
condition requires that $k(1-g_k)$ is constant for all groups of
connectivity $k$ at the threshold, implying that $g_k\sim 1 - 1/ k
\lambda$; i.e., a larger portion of individuals must be immunized in
groups with larger connectivity.  In this scheme the total density of
immunized individuals can be easily calculated by averaging $g_k$ over
the various connectivity classes.  The fraction of non-immunized
individuals $1-g_k$ can not be larger than one, thus we focus only on
classes with connectivity such that the reproductive number $k >
\lambda^{-1}$.  To eradicate the infection, we need that $g_k \geq 1-
1/k\lambda$ in all classes with connectivity $k>\lambda^{-1}$,
defining the critical fraction of immunized individuals as
\begin{equation}
g_c= \sum_{k>\lambda^{-1}}(1- \frac{1}{\lambda k})P(k).
\end{equation}
In order to perform an explicit calculation for the BA model, we use again
the continuous $k$ approximation  \cite{barab99}.  In this case we obtain that
\begin{equation}
g_c= \frac{1}{3}(m\lambda)^2.
\end{equation}
This result can be readily extended to SF networks with arbitrary
$\gamma$ values, and it is worth remarking that this recipe is along
the lines of that introduced in the immunization of heterogeneously
populated groups \cite{anderson92}. Recently, a similar strategy has
been put forward in Ref.~\cite{aidsbar} by proposing to cure  
with proportionally higher rates the most connected nodes.

\begin{figure*}[t] 
  \centerline{\epsfig{file=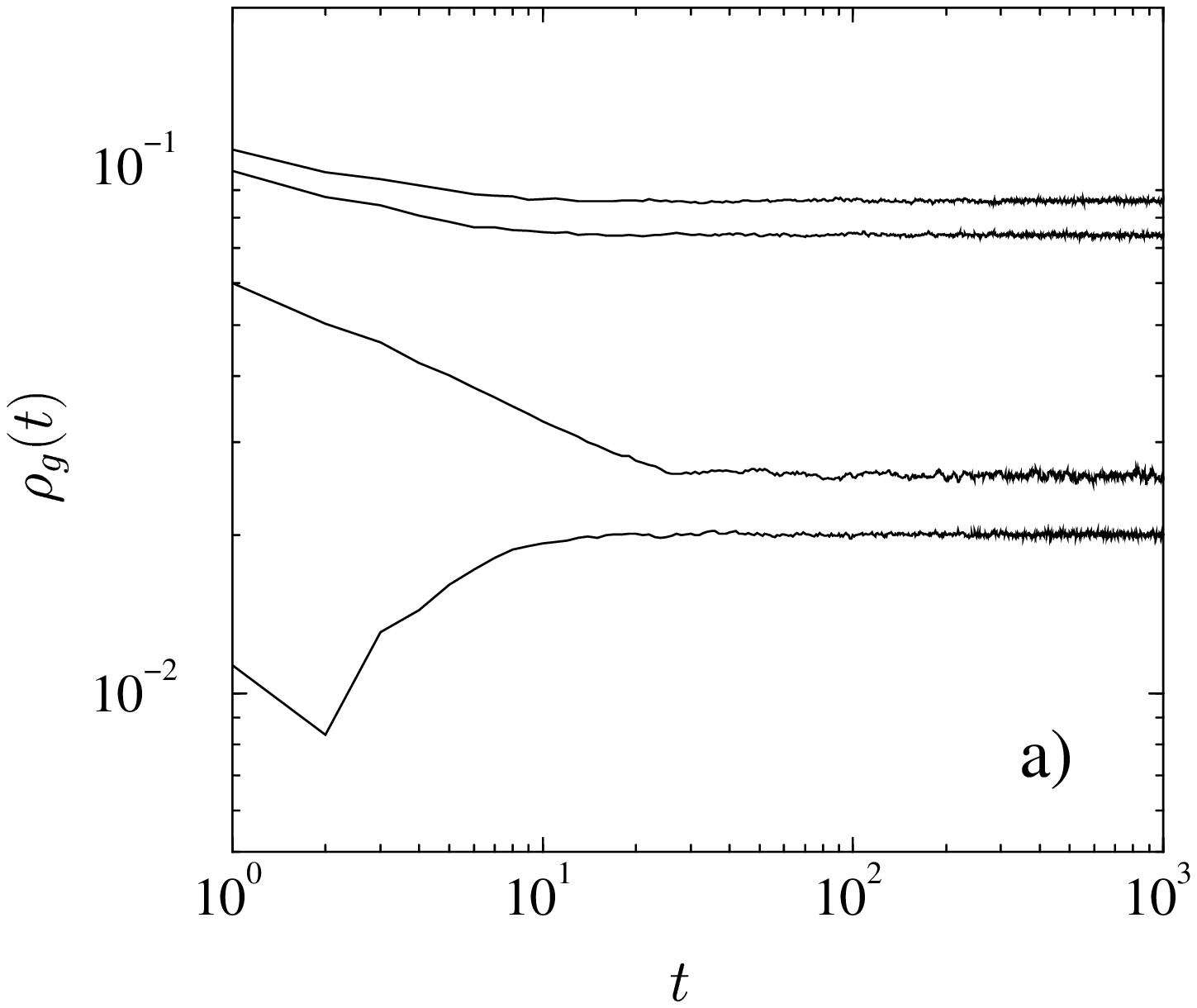, width=8cm}
    \epsfig{file=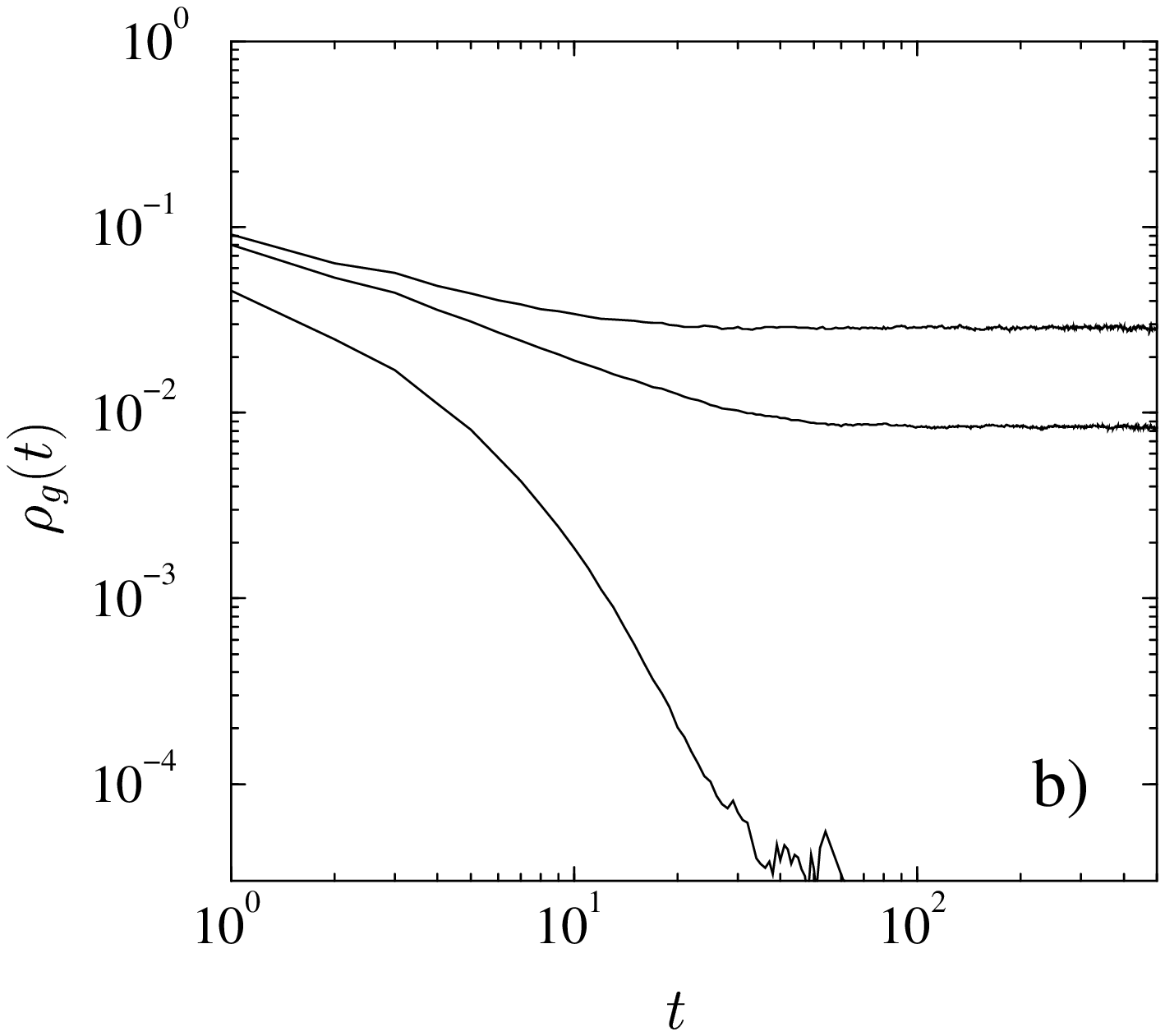, width=8cm}}

  \caption{a) Typical plots of $\rho_g(t)$ as a function of time,
    averaged over $100$ starting configurations, from computer
    simulations of the SIS model in BA networks with uniform
    immunization for different values of $g$. From top to bottom:
    $g=0.1$, $0.14$, $0.3$, and $0.5$. For all values of $g$ shown,
    the endemic state is reached in a sufficiently large network. %
    ~b)Typical plots of $\rho_g(t)$ as a function of time, averaged over
    $100$ starting configurations, for the SIS model in BA networks
    with targeted immunization for different values of $g$.  From top
    to bottom: $g=0.1$, $0.14$, and $0.3$.  For the last value, larger
    than the critical immunization, all runs die for any network
    size.}

\end{figure*}

\subsection{Targeted immunization}

While proportional immunization schemes are effective in finally
introducing an well-defined immunization threshold, the very peculiar
nature of SF networks allows to define more efficient strategies based
on the nodes' hierarchy.  In particular, it has been shown that SF
networks posses a noticeable resilience to random connection
failures~\cite{barabasi00,newman00,havlin01}, which implies that the
network can resist a high level of damage (disconnected links),
without loosing its global connectivity properties; i.e., the
possibility to find a connected path between almost any two nodes in
the system.  At the same time, SF networks are strongly affected by
selective damage; if a few of the most connected nodes are removed,
the network suffers a dramatic reduction of its ability to carry
information \cite{barabasi00,newman00,havlin01}.  Applying this
argument to the case of epidemic spreading, we can devise a {\em
  targeted} immunization scheme in which we progressively make immune
the most highly connected nodes, i.e., the ones more likely to spread
the disease. While this strategy is the simplest solution to the
optimal immunization problem in heterogeneous populations
\cite{anderson92}, its efficiency is comparable to the uniform
strategies in networks with finite connectivity variance. In SF
networks, on the contrary, it produces an arresting increase of the
network tolerance to infections at the price of a tiny fraction of
immune individuals.

Let us consider the situation in which a fraction $g$ of the
individuals with the highest connectivity are successfully immunized.
This corresponds, in the limit of a large network, to 
the introduction an upper threshold $k_t$,
such that all nodes with connectivity $k > k_t$ are immune. The
fraction of immunized individuals is then given by
\begin{equation}
g= \sum_{k>k_t}P(k),
\end{equation}
relation that renders $k_t$ an implicit function of $g$.
The presence of the cut-off $k_t(g)$ defines the new average
quantities $\left <k \right>_t=\sum_m^{k_t}kP(k)$ and $\left< k^2
\right>_t=\sum_m^{k_t}k^2P(k)$, which are on their turn function of $g$.
At the same time, all links emanating from immunized individuals can
be considered as if they were removed. The probability $p(g)$ that any
link will lead to an immunized individual is then given by
\begin{equation}
p(g)=\frac{\sum_{k>k_t(g)}kP(k)}{\sum_k kP(k)},
\end{equation}
and if we consider that this fraction $p(g)$ of links are effectively
removed, the new connectivity distribution after the immunization of a
fraction $g$ of the most connected individuals is \cite{havlin01}:
\begin{equation}
P_g(k)=\sum_{q \geq k}^{k_t} P(q) \binom{q}{k}  (1-p)^kp^{q-k}.
\end{equation}
The new distribution (after cut-off introduction and link removal)
yields the first two moments $\left<k\right>_g=\left<k\right>_t(1-p)$
and $\left<k^2\right>_g=\left<k^2\right>_t(1-p)^2+\left<k\right>_t
p(1-p)$ \cite{havlin01}.  By recalling Eq.~(\ref{thr}), the critical
fraction $g_c$ of immune individuals needed to eradicate the infection
will be given by the relation
\begin{equation}
  \frac{\left<k^2\right>_{g_c}}{\left<k\right>_{g_c}}\equiv
  \frac{\left<k^2\right>_t}{\left<k\right>_t}(1-p(g_c)) 
  +p(g_c) =  \lambda^{-1}.
\label{th_targ}
\end{equation}
An explicit calculation for the BA network
in the continuous $k$ approximation  yields that the density 
of immunized nodes is related to the connectivity threshold as 
\begin{equation}
g=1-\int_m^{k_t}P(k)dk=m^2k_t^{-2}.
\end{equation}
By inverting this relation we obtain that the connectivity threshold 
is $k_t=mg^{-1/2}$, yielding that
\begin{equation}
  p(g)=\frac{1}{2m} \left( 1-\int_m^{k_t}kP(k)dk \right) = g^{1/2}.
\end{equation}
As well, we can obtain 
$\left< k \right>_t\simeq 2m$ and 
 $\left< k^2
\right>_t\simeq 2m^2\ln(g^{-1/2})$ as  $k_t=mg^{-1/2}\to\infty$.
By inserting these values into  Eq.~(\ref{th_targ}) we obtain 
the approximate solution for the immunization threshold in the case of 
targeted immunization as  
\begin{equation}
g_c\simeq \exp(-2/m\lambda).
\label{eq:gcBA}
\end{equation}
This clearly indicates that the targeted immunization program is
extremely convenient in SF networks where the critical immunization is
exponentially small in a wide range of spreading rates $\lambda$.
Also in this case, the present result can be generalized for SF
networks with  arbitrary
connectivity exponent $\gamma$.

In order to test the targeted immunization scheme we have implemented
numerical simulations of the SIS model on the WS and BA networks by
immunizing the $g N$ nodes with the highest connectivity.  Note that,
for a given network, this method is essentially deterministic: Once we
identify the hierarchy in the node's connectivity distribution, we
proceed to protect those nodes on top of the list. Simulations are
performed at a fixed spreading rate $\lambda=0.25$.  In Fig. 1(a) we
report the behavior of the prevalence of infected nodes for the WS
network with targeted immunization; the results corresponding to the
BA graph are plotted in Fig. 2(a). In the case of the WS network, the
behavior of the prevalence as a function of $g$ is equivalent in the
uniform and targeted immunization procedures. The connectivity
fluctuations are small, and the immunization of the most connected
nodes is equivalent to the random choice of immune nodes.  This
confirms that targeted strategies do not have a particular efficiency
in systems with limited heterogeneity.  On the contrary, in the case
of the BA network, we observe a drastic variation in the prevalence
behavior. In particular, the prevalence suffers a very sharp drop and
exhibits the onset of an immunization threshold above which no endemic
state is possible (zero infected individuals). A linear extrapolation
from the largest values of $g$ yields an estimate of the very
convenient threshold $g_c \simeq 0.16$. 
This definitely shows that SF networks are
highly sensitive to the targeted immunization of a small fraction of
the most connected nodes (see Fig. 2(a) and Fig. 3(b)). While these networks are
particularly weak in face of infections, the good news consist in the
possibility to devise immunization strategies which are extremely
effective.

\section{Discussion and conclusions}

The present results indicate that the SF networks' susceptibility to
epidemic spreading is reflected also in an intrinsic difficulty in
protecting them with local---uniform---immunization. On a global
level, uniform immunization policies are not satisfactory and, in
analogy with disease spreading in heterogenous media, only targeted
immunization procedures achieve the desired lowering of epidemic
outbreaks and prevalence. This evidence radically changes the usual
perspective of the regular epidemiological framework. Spreading of
infectious or polluting agents on SF networks, such as food or social
webs, might be contrasted only by a careful choice of the immunization
procedure. In particular, these procedures should rely on the
identification of the most connected individuals. The protection of
just a tiny fraction of these individuals raises dramatically the
tolerance to infections of the whole population

\begin{figure}[t]
  \centerline{\epsfig{file=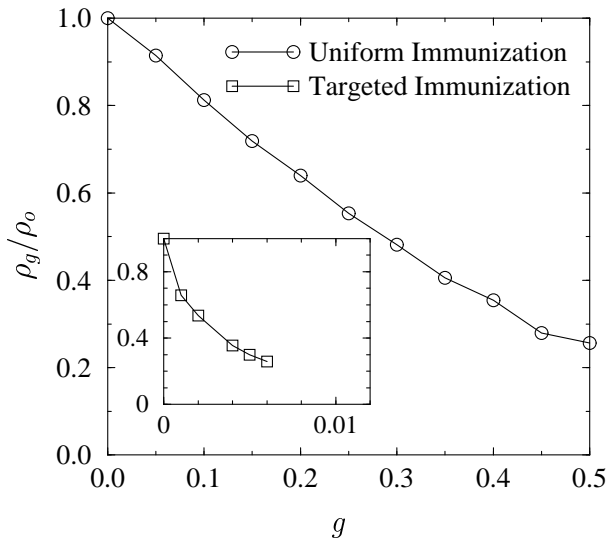, width=8cm}} 

  \caption{Reduced prevalence $\rho_g/ \rho_0$ from computer
    simulations of the SIS model in a portion of a real Internet map
    with uniform (main plot) and targeted (inset) immunization at a
    fixed spreading rate $\lambda=0.25$.  We only consider values of
    the immunization for which almost all the runs survive up to the
    end.  This explains the short range of values of $g$ shown for the
    targeted immunization case.}

\end{figure}

A practical example is provided by the spreading of viruses in the
Internet \cite{ieee93}.  The SF nature of this network is the outcome
of a connectivity redundancy which is very welcome because it ensures
a greater error tolerance than in less connected networks. On the
other hand, despite the large use of anti-virus software which is
available on the market within days or weeks after the first virus
incident report, the average lifetime of digital epidemics is
impressively large (10-14 months) \cite{pv01a}.  Numerical simulations
of the SIS model on real maps of the Internet can provide further
support to our picture. The SIS model is, in fact, well suited to
describe DNS-cache computer viruses~\cite{bellovin} (the so-called
``natural computer viruses''), and different digital viruses can be
modeled by considering the random neighbor version of the
model~\cite{marro99}); i.e. infected e-mails can be sent to different
nodes which are not nearest neighbors.  The map considered here,
provided by the National Laboratory for Applied Network Research
(NLANR) and available at the web site
http://www.moat.nlanr.net\-/Rou\-ting/rawdata/, contains $6313$ nodes
and $12362$ links, corresponding to an average connectivity
$\left<k\right> = 3.92$. The connectivity distribution is scale-free,
with a characteristic exponent $\gamma \simeq 2.2$ \cite{alexei}.  Our
simulations are performed at a fixed spreading rate $\lambda=0.25$,
averaging over at least $2500$ different starting configurations.  We
implement both the uniform and the targeted immunization procedures.
The results obtained clearly indicate that the behavior is completely
analogous to that found on the BA network.  Fig.~4 illustrates that,
while uniform immunization does not allow any drastic reduction of the
infection prevalence---the immunization of $25\%$ of the nodes reduces
by less than a factor $1/2$ the relative prevalence---the targeted
immunization drastically removes the occurrence of endemic states even
at very low value of the immunization parameter. The fact that SF
networks can be properly secured only by a selective immunization,
points out that an optimized immunization of the Internet can be
reached only through a global immunization organization that secures a
small set of selected high-traffic routers or Internet domains.
Unfortunately, the self-organized nature of the Internet does not
allow to easily figure out how such an organization should operate.

The present results also appear to have potentially interesting
implications in the case of human sexual disease control
\cite{anderson92,lloyd01}.  Most sexually transmitted diseases cannot be
characterized without including the noticeable differences of sexual
activity within a given population. Epidemic modeling is thus based on
partitioning population groups by the number of sexual partners per
unit time \cite{anderson92}.  This implicitly
corresponds to the knowledge of the probability distribution function
$P(k)$ that gives the fraction of the population within the $k$ class.
The recent observation that the web of human sexual contacts exhibits
scale-free features \cite{amaral01} points out that also sexually
transmitted diseases are eventually spreading in a network with
virtually infinite heterogeneity. It follows that concepts such as the
mean number of sexual partners or its variance are not good indicators
in this case. As well, the definition of a core group of
``super-spreader'' individuals could be a non well-defined concept
because of the lack of precisely defined thresholds or characteristic
magnitudes in the scale-free distribution of sexual contacts.
Nevertheless, the striking effectiveness of targeted
immunization indicates that control and prevention campaigns should be
strongly focused at the most promiscuous individuals.  These represent
the most connected nodes of the network and are thus the key
individuals in the spreading of the infection.

While the simple SIS model is very instructive, many other ingredients
should be considered in a more realistic representation of real
epidemics \cite{anderson92,epidemics}.  One would also want to add
simple rules defining the temporal patterns of networks such as the
frequency of forming new connections, the actual length of time that a
connection exists, or different types of connections.  These dynamical
features are highly valuable experimental inputs which are necessary
ingredients in the use of complex networks theory in epidemic
modeling.

\begin{acknowledgments}
  
  This work has been partially supported by the European Network
  Contract No. ERBFMRXCT980183.  The NLANR project is supported by
  NSF. RP-S also acknowledges support from the grant CICYT PB97-0693.
  We thank A. Barrat, M. Mezard, M.-C.  Miguel, R. V.  Sol{\'e}, S.
  Visintin, and R.  Zecchina for helpful comments and discussions. We
  are grateful to A.~L. Lloyd and R.~M.~May for enlightening
  suggestions and for pointing out to us some fundamental references.

\end{acknowledgments}

\end{document}